*On Fresnel-Airy Equations, Fabry-Perot Resonances and Surface Electromagnetic Waves in Arbitrary Bianisotropic Metamaterials, including with Multi-Hyperbolic Fresnel Wave Surfaces.*

Maxim Durach*, Felix Williamson, Jacob Adams, Tonilynn Holtz, Pooja Bhatt, Rebecka Moreno, Franchescia Smith

Department of Physics and Astronomy, Georgia Southern University, Statesboro, Georgia 30461, USA

Corresponding author: *mdurach@georgiasouthern.edu

**Abstract**

We introduce a theory of optical responses of bianisotropic layers with arbitrary effective medium parameters, which results in generalized Fresnel-Airy equations for reflection and transmission coefficients at all incidence directions and polarizations. The poles of these equations provide explicit expressions for the dispersion of Fabry-Perot resonances and surface electromatic waves in bianisotropic layers and interfaces. The existence conditions of these resonances are topologically related to the zeros of the high-k characteristic function $h(\boldsymbol{k}) = 0$ of bulk bianisotropic materials and Durach et al. taxonomy of bianisotropic media according to the hyperbolic topological classes [Applied Sciences, 10(3), 763 (2020); Optics Communications, 476, 126349 (2020)].

1. Introduction

The studies of bianisotropic optical materials (bianisotropics [1-3]) emerged [4-7] from the investigations of magnetoelectric effects in moving media first observed by Roentgen and Wilson [4,8-10] and from consideration of magnetoelectric crystals started by Landau, Lifshitz, and Dzyaloshinskii [11-15]. Through the machinery of transformational optics [16-18], it has been demonstrated that light propagation through bianisotropic optical metamaterials can serve as simulations of Universe-aged cosmology problems from times immemorial [19,20]. Now bianisotropic materials attract a lot of attention due to the practical development of complex anisotropic and bianisotropic photonic metamaterials in a wide range of frequency bands [21,22]. Metamaterial structures feature many exotic optical properties important for applications, such as superresolution imaging and sensing, optical cloaking, subwavelength optical confinement and guiding, emission rate and directivity control, optical magnetism etc. [21-24] Typically, these properties depend on metamaterials being anisotropic and bianisotropic optical media [21,22].

Bianisotropic media are the most general case of local linear media [3-7,25] and are characterized by the effective material parameters grouped into a 6x6 material parameters matrix $\widehat{M}$, which relate the electric displacement field $\boldsymbol{D}$ and magnetic field $\boldsymbol{B}$ with fields $\boldsymbol{E}$ and $\boldsymbol{H}$:

$$\begin{pmatrix} \boldsymbol{D} \\ \boldsymbol{B} \end{pmatrix} = \widehat{M} \begin{pmatrix} \boldsymbol{E} \\ \boldsymbol{H} \end{pmatrix} = \begin{pmatrix} \hat{\epsilon} & \hat{X} \\ \hat{Y} & \hat{\mu} \end{pmatrix} \begin{pmatrix} \boldsymbol{E} \\ \boldsymbol{H} \end{pmatrix}. \qquad (1)$$

The 3x3 matrices $\hat{\epsilon}, \hat{\mu}, \hat{X}, \hat{Y}$ are dielectric permittivity, magnetic permeability and two magnetoelectric coupling matrices respectively. Until recently very few general concrete properties of bulk bianisotropic media were established beyond formulaic apparatus due to the complexity of these materials and multiparametric nature of these media [3,25-27]. Bulk bianisotropic materials feature quartic isofrequency Fresnel wave surfaces [28-31] which can be described using the direction-dependent index of refraction operator theory [32,33]. Note that, unlike quadric and cubic surfaces, quartic surfaces have not been completely explored and fully classified in mathematical sciences [34,35]. In 2020 Durach et al. proposed a taxonomy of bianisotropic optical media based on the number of double cones that isofrequency Fresnel wave surfaces feature at high k-vectors [32,33]. According to this taxonomy, both bianisotropic and anisotropic media can be classified by 5 topological classes non-, mono-, bi-, tri-, and tetrahyperbolic materials [32,33,36-40]. An example of a tetrahyperbolic isofrequency Fresnel wave surface is shown in Fig. 1(a) for the effective medium with matrix $\widehat{M}$ color coded in the bottom of that panel. Durach et al. taxonomy [32,33] provides a framework for future advances not only in the field of bianisotropic metamaterials, but also for hyperbolic metamaterials, already known for applications in optical imaging, hyperlensing, and emission rate and directivity control utilizing the diverging optical density of high-k states [22,24,39,40].

Researchers in bianisotropics are attempting to use the optical responses of specific bianisotropic metamaterial and metasurface structures including reflection, transmission, and scattering characteristics, resonances, surface electromagnetic waves (SEWs), and various properties of field distributions to retrieve the information about the effective material parameters [41-54] and the constituent meta-atoms [55-58]. Another important direction is to homogenize metamaterials composed of specific meta-atoms into an effective medium [59-75] or inversely to use structure synthesis or modular approach of materiatronics [76-81] to decompose the effective medium into constituent meta-atoms. At the same time, predicting and classifying the general features of the optical responses and electromagnetic fields in bianisotropic effective media is challenging and limited to special important cases [82-102], and has not been completed for a generic bianisotropic media due to the difficulty in explicit determination of the waves propagating and decaying in a particular direction in such a medium. For example, obtaining Fresnel equations can be challenging at generic bianisotropic interfaces [103]. Another reason for the complications is the need to establish the additional boundary conditions (ABCs) in extreme non-reciprocal cases [104]. Even though a classification of individual SEWs at bianisotropic boundaries can be introduced based on propagation and penetration characteristics of the waves [105] and the invariance classes for individual SEWs with respect to variations of material parameters can be established [105], the classification of the isofrequency curves for SEWs is difficult to obtain [106-111].

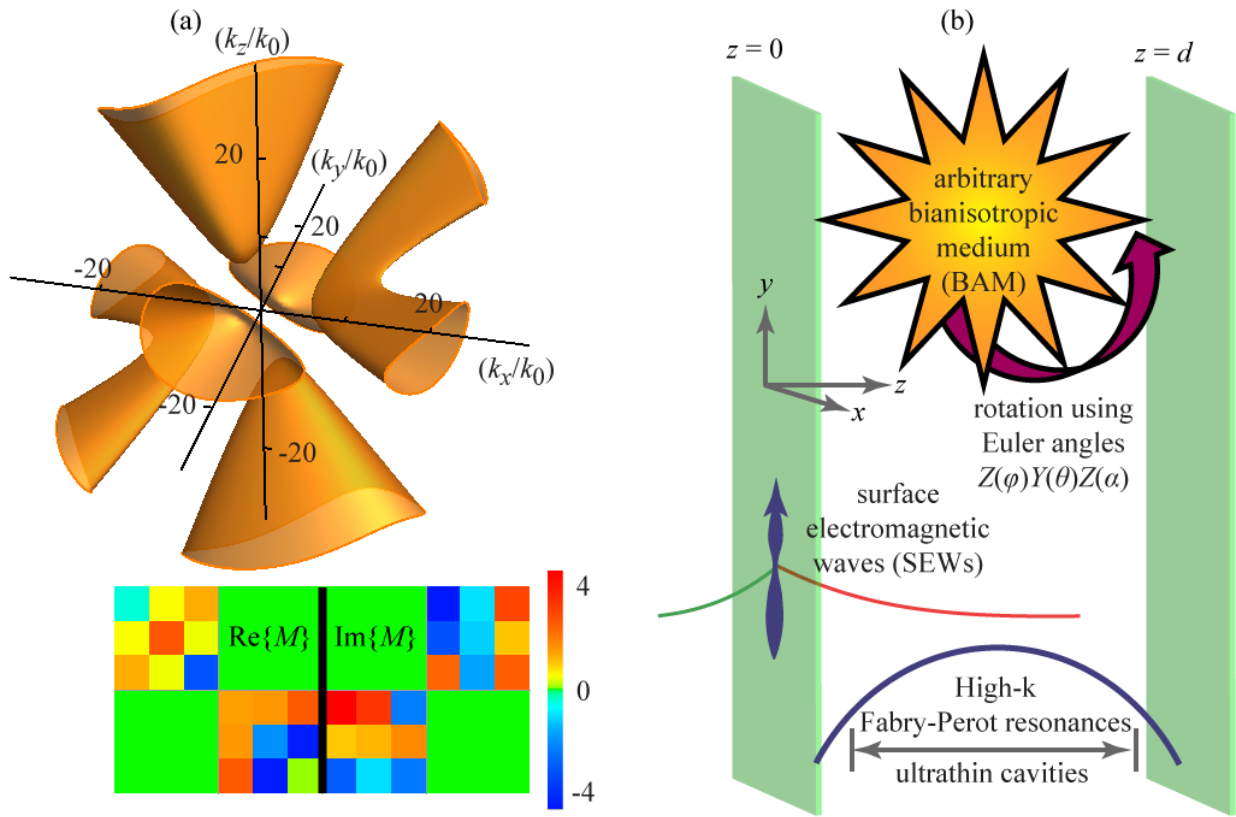

Fig. 1. (a) Isofrequency Fresnel wave surface in k-space for a tetrahyperbolic bianisotropic material with $\widehat{M}$ matrix illustrated with color coding in the bottom of the panel. (b) Schematic illustrating the conventions in this paper as described in the text.

In general, finding resonances and SEWs in stratified bianisotropic arrays represents a transcendental optics problem and their general properties have not been established previously in the literature [112-117]. The optical responses of an isotropic layer are described by Airy formulas [118,119], whose poles correspond to the Fabry-Perot resonances [120,121] and describe both guided and SEW modes of the layers [122,123]. In isotropic materials and some bianisotropic materials, the electromagnetic fields are characterized by specific polarizations [124-128], but generally the anisotropic and bianisotropic layers and metasurfaces feature multimode resonances with coupled polarizations, which has been an additional complication [129-131].

In this paper we introduce a complete theory of optical responses of arbitrary bianisotropic planar cavities and wave guides. We obtain the explicit expressions for the reflection and transmission coefficients in any polarization from both sides of a generic bianisotropic layer, and an explicit expression which describes the Fabry-Perot resonances and surface electromagnetic waves propagating in such bianisotropic layers.

We relate the conditions for these resonances with the orientation of the bulk Fresnel wave surface with respect to the boundaries of the layer (see Fig. 1 for schematic). The regions where Fabry-Perot resonances may propagate are separated from the regions of surface electromagnetic waves by the zero curve of the high-k characteristic function, $h(\mathbf{k})$, characterizing the topological phase of bulk bianisotropic materials according to Durach et al. taxonomy [32,33].

We demonstrate that thin subwavelength layers form cavities supporting high-k Fabry-Perot resonances if the bianisotropic material is oriented such that the high-k direction, given by the zero of high-k characteristic function, is close to perpendicular to the interfaces of the layer. The density of Fabry-Perot resonances diverges as the bianisotropic material direction approaches $h(\mathbf{k}) = 0$. As the thickness of the layer increases, the resonance spectrum becomes more populated with lower-k modes of the bulk bianisotropic media. The SEWs appear as arcs in the regions where evanescent waves are supported between the regions with propagating waves. The resonances of the layer are represented as the interference of multiple channels of energy transfer through the cavity.

2. Methods

*Theory of Optical Response and Resonances of Bianisotropic Layers*

We base our theory of optical responses and resonances of bianisotropic layers in the characteristic matrix formalism [132,33]. The explicit expression for the 4x4 characteristic matrix of bianisotropic media $\widehat{\Delta}$ was recently obtained in terms of the material parameters contained within $\widehat{M}$ and the parallel to the layer component of the k-vector $\boldsymbol{k}_\parallel = k_0(q_x, q_y)$ [33]. For $\boldsymbol{k}_\parallel = 0$ the characteristic matrix $\widehat{\Delta}$ corresponds to the index of refraction operator $\widehat{N}$ in the z-direction [32,33]. The eigenvalues $q_z^{(i)} = q_z^{(i)}(q_x, q_y)$, $i = 1,..,4$ of the characteristic matrix $\widehat{\Delta}$ provide the shape of the isofrequency Fresnel wave surface. Electromagnetic fields in a bianisotropic media can be expressed in terms of the eigenvalues $q_z^{(i)}$ and eigenvectors $\Gamma_\parallel^{(i)} = \left(E_x^{(i)}, E_y^{(i)}, H_x^{(i)}, H_y^{(i)}\right)^T$ of the matrix $\widehat{\Delta}$ using eigenvalue decomposition

$$\widehat{\Delta} = \widehat{F} \cdot \widehat{Q} \cdot \widehat{F}^{-1}, \tag{2}$$

where $\widehat{F} = \left(\Gamma_\parallel^{(1)}, \Gamma_\parallel^{(2)}, \Gamma_\parallel^{(3)}, \Gamma_\parallel^{(4)}\right)$ is the eigenvector matrix composed of eigenvectors of matrix $\widehat{\Delta}$, and $\widehat{Q} = \text{diag}\left\{q_z^{(1)}, q_z^{(2)}, q_z^{(3)}, q_z^{(4)}\right\}$.

The z-dependence of the x- and y-components of the field can be expressed as

$$\begin{pmatrix} E_x \\ E_y \\ H_x \\ H_y \end{pmatrix} = \begin{pmatrix} \sum A_i E_x^{(i)} e^{ik_0 q_z^{(i)} z} \\ \sum A_i E_y^{(i)} e^{ik_0 q_z^{(i)} z} \\ \sum A_i H_x^{(i)} e^{ik_0 q_z^{(i)} z} \\ \sum A_i H_y^{(i)} e^{ik_0 q_z^{(i)} z} \end{pmatrix} = \widehat{F} \cdot \exp(ik_0 \widehat{Q} z) \cdot A, \tag{3}$$

where the amplitude vector $A = \left(A^{(1)}, A^{(2)}, A^{(3)}, A^{(4)}\right)^T$ is composed of the amplitudes of the eigenmodes in Eq. (3).

For a wave with $\boldsymbol{k}_\parallel = k_0(q_x, 0)$ in vacuum $\widehat{Q}_0 = \text{diag}\{q_z, q_z, -q_z, -q_z\}$ is the eigenvalue matrix with $q_z = \sqrt{1 - q_x^2}$. Two sets of eigenvectors in vacuum $\widehat{F}_0 = \left(\Gamma_\parallel^{(1)}, \Gamma_\parallel^{(2)}, \Gamma_\parallel^{(3)}, \Gamma_\parallel^{(4)}\right)$ are typically introduced. First, for linear polarization we use

$$\Gamma_{\parallel,\text{TM}}^{(1)} = \begin{pmatrix} E_x \\ E_y \\ H_x \\ H_y \end{pmatrix} = \begin{pmatrix} q_z \\ 0 \\ 0 \\ 1 \end{pmatrix}, \Gamma_{\parallel,\text{TE}}^{(2)} = \begin{pmatrix} 0 \\ 1 \\ -q_z \\ 0 \end{pmatrix}, \Gamma_{\parallel,\text{TM}}^{(3)} = \begin{pmatrix} -q_z \\ 0 \\ 0 \\ 1 \end{pmatrix}, \Gamma_{\parallel,\text{TE}}^{(4)} = \begin{pmatrix} 0 \\ 1 \\ q_z \\ 0 \end{pmatrix}. \tag{4}$$

Secondly, the left- and right-circularly polarized eigenvectors $\Gamma_{\parallel,L,R}^{(i)}$ are

$$\Gamma_{\parallel,L}^{(1)} = \Gamma_{\parallel,\text{TM}}^{(1)} + i\Gamma_{\parallel,\text{TE}}^{(2)}, \Gamma_{\parallel,R}^{(2)} = \Gamma_{\parallel,\text{TE}}^{(2)} + i\Gamma_{\parallel,\text{TM}}^{(1)}, \tag{5a}$$

$$\Gamma^{(3)}_{\|,L} = \Gamma^{(3)}_{\|,TM} + i\Gamma^{(4)}_{\|,TE}, \Gamma^{(4)}_{\|,R} = \Gamma^{(4)}_{\|,TE} + i\Gamma^{(3)}_{\|,TM}. \tag{5b}$$

We consider a bianisotropic layer placed into the vacuum at $0 < z < d$ (Fig.1(b)). First, consider the incidence from the $z < 0$ side of TM, or left-handed circular polarized $\Gamma^{(1)}_\|$ waves depending on the basis eigenvectors chosen (Eqs. (4)-(5)). The z-dependence of the fields $V_-(z)$ in the vacuum at $z < 0$ is

$$V_-(z) = \Gamma^{(1)}_\| e^{ik_z z} + \left(r_{13}\Gamma^{(3)}_\| + r_{14}\Gamma^{(4)}_\|\right) e^{-ik_z z}, \tag{6}$$

where $k_z = k_0 q_z$. On the other side of the layer at $z > d$

$$V_+(z) = \left(t_{11}\Gamma^{(1)}_\| + t_{12}\Gamma^{(2)}_\|\right) e^{ik_z(z-d)} \tag{7}$$

Similarly, the functions $V_-(z)$ and $V_+(z)$ can be written for TM and TE (or left- and right-handed) polarization incidence from both sides of the bianisotropic layer, with different reflection and transmission coefficients $r_{ab}, t_{ab}$, where $a = 1,..,4$ corresponds to the number of incidence wave and $b = 1,...,4$ denotes the number of the reflection or transmission wave in the desired basis. We denote the eigenvalues and eigenvectors in the bianisotropic layer as $\hat{Q}_B = \text{diag}\{q^{(1)}_z, q^{(2)}_z, q^{(3)}_z, q^{(4)}_z\}$ and $\hat{F}_B = \left(\Gamma^{(1)}_{\|,B}, \Gamma^{(2)}_{\|,B}, \Gamma^{(3)}_{\|,B}, \Gamma^{(4)}_{\|,B}\right)$, respectively. The amplitudes for the fields in the bianisotropic layer Eq. (3) are denoted as $A_a$ and grouped into vector $A$, where the subscript "$a$" depends on the incidence wave (see Eq. (3)). For all types of incidence, the boundary conditions can be written as

$$V_-(0) = \hat{F}_B \cdot A, \tag{8}$$

$$\hat{F}_B \cdot \exp(ik_0 d\hat{Q}) \cdot A = V_+(d). \tag{9}$$

The fields to the right and to the left of the layer can be related using Eqs. (8)-(9) and the coefficients $A$ can be obtained as

$$V_+(d) = \hat{F}_B \cdot \exp(ik_0 d\hat{Q}) \cdot \hat{F}_B^{-1} \cdot V_-(0) = \exp(ik_0 d\hat{\Delta}) \cdot V_-(0) \tag{10}$$

$$A = \hat{F}_B^{-1} \cdot V_-(0) \tag{11}$$

Eq. (10) can be written for all 4 types of excitation simultaneously in the matrix form

$$\hat{F}_{0-} + \hat{F}_{0+} \cdot R = \exp(ik_0 d\hat{\Delta}) \cdot (\hat{F}_{0+} + \hat{F}_{0-} \cdot R) \tag{12}$$

using the *optical response matrix* $\hat{R}$

$$\hat{R} = \begin{pmatrix} t_{11} & t_{21} & r_{31} & r_{41} \\ t_{12} & t_{22} & r_{32} & r_{42} \\ r_{13} & r_{23} & t_{33} & t_{43} \\ r_{14} & r_{24} & t_{34} & t_{44} \end{pmatrix}, \qquad (13)$$

where we group the waves with positive $S_z > 0$ and negative $S_z < 0$ z-components of the Poynting vector into matrices $\hat{F}_{0+}$ and $\hat{F}_{0-}$, respectively

$$\hat{F}_{0+} = \left(\Gamma_\parallel^{(1)}, \Gamma_\parallel^{(2)}, 0, 0\right), \quad \hat{F}_{0-} = \left(0, 0, \Gamma_\parallel^{(3)}, \Gamma_\parallel^{(4)}\right). \qquad (14)$$

From Eq. (12) we provide the formal solution to the problem of finding the response matrix $\hat{R}$ of an arbitrary bianisotropic layer

$$\hat{R} = \left(\hat{F}_{0+} - \exp(ik_0 d\hat{\Delta}) \cdot \hat{F}_{0-}\right)^{-1} \left(\exp(ik_0 d\hat{\Delta}) \cdot \hat{F}_{0+} - \hat{F}_{0-}\right) \qquad (15)$$

which is composed of all possible reflection and transmission coefficients $r_{ab}, t_{ab}$. Our result Eq. (15) generalizes the Airy formula for isotropic layers [118,119] to arbitrary bianisotropic media. The poles of Eq. (15) describe Fabry-Perot resonances and surface electromagnetic waves in bianisotropic media as described below.

Note that the response matrix in the form $\hat{R} = \hat{1} \cdot e^{ik_z d}$ corresponds to the complete invisibility (nihility) of the layer in both amplitude and phase in all polarizations, while deviations from the identity matrix in the response matrix corresponds to presence of forward or backward scattering [133]. Comparing Eqs. (5) and (10)-(15) for linear and circular polarizations we see that response matrices in linear and circular polarization are related as

$$R_{circ} = (1 + i\hat{B})^{-1} R_{lin}(1 + i\hat{B}), \qquad \hat{B} = \begin{pmatrix} 0 & 1 & 0 & 0 \\ 1 & 0 & 0 & 0 \\ 0 & 0 & 0 & 1 \\ 0 & 0 & 1 & 0 \end{pmatrix}$$

The formalism of the optical response matrix $\hat{R}$ can be used for the design of novel bianisotropic planar optical components, such as cavities [130] or wave-plates [131]. Below we investigate Fabry-Perot resonances and surface electromagnetic waves using this formalism. The general condition for Fabry-Perot resonances and SEWs in bianisotropic media can be found as poles of the generalized Airy formula given by Eq. (15), or equivalently as zeros of the denominators $\delta_A$:

$$\delta_A = \det\{\hat{F}_{0+} - \exp(ik_0 d\hat{\Delta}) \cdot \hat{F}_{0-}\} \qquad (16)$$

After tedious algebra Eq. (16) can be rewritten as

$$\delta_A = \left(D_{12}^- D_{34}^+ e^{i\varphi_{12}} + D_{13}^- D_{24}^+ e^{i\varphi_{13}} + D_{14}^- D_{23}^+ e^{i\varphi_{14}} + D_{23}^- D_{14}^+ e^{i\varphi_{23}} + D_{24}^- D_{13}^+ e^{i\varphi_{24}} + D_{34}^- D_{12}^+ e^{i\varphi_{34}}\right)/D_B \qquad (17)$$

where $\varphi_{ij} = ik_0\left(q_z^{(i)} + q_z^{(j)}\right)d$, $D_{ij}^- = \det\left(\Gamma_\parallel^{(1)}, \Gamma_\parallel^{(2)}, -\Gamma_{\parallel,B}^{(i)}, -\Gamma_{\parallel,B}^{(j)}\right)$, and $D_{pq}^+ = \det\left(\Gamma_{\parallel,B}^{(p)}, \Gamma_{\parallel,B}^{(q)}, -\Gamma_\parallel^{(3)}, -\Gamma_\parallel^{(4)}\right)$, $D_B = \det(\hat{F}_B)$.

***Optical Response of Bianisotropic Layers in Terms of Fresnel Coefficients at Bianisotropic Surfaces.***

To provide an insight into the physics of the resonance condition of Eqs. (16)-(17) we derive the Fresnel coefficients for a generic boundary of bianisotropic materials and express Eq. (15) in terms of these coefficients as is frequently done for isotropic layers. To accomplish this, we need to assume that there are no additional waves on the boundary, otherwise additional boundary conditions (ABCs) introduced in Ref. [104] are needed. We assume that waves $\Gamma_{\parallel,B}^{(1)}, \Gamma_{\parallel,B}^{(2)}$ have positive $S_z > 0$ and $\Gamma_{\parallel,B}^{(3)}, \Gamma_{\parallel,B}^{(4)}$ have negative $S_z < 0$. Considering the incidence of $\Gamma_{\parallel,B}^{(1)}, \Gamma_{\parallel,B}^{(2)}$ from the $z < 0$ side the Fresnel coefficients $r_{ab}^F, t_{ab}^F$ can be obtained from boundary conditions [104,105] and are given by

$$\begin{pmatrix}\hat{T}_{12}\\ \hat{R}_{12}\end{pmatrix} = \left(\Gamma_\parallel^{(1)}, \Gamma_\parallel^{(2)}, -\Gamma_{\parallel,B}^{(3)}, -\Gamma_{\parallel,B}^{(4)}\right)^{-1} \cdot \left(\Gamma_{\parallel,B}^{(1)}, \Gamma_{\parallel,B}^{(2)}\right). \qquad (18)$$

For the incidence of $\Gamma_{\parallel,B}^{(3)}, \Gamma_{\parallel,B}^{(4)}$ from the $z > d$ side, the Fresnel coefficients are found from

$$\begin{pmatrix}\hat{T}_{34}\\ \hat{R}_{34}\end{pmatrix} = \left(\Gamma_\parallel^{(3)}, \Gamma_\parallel^{(4)}, -\Gamma_{\parallel,B}^{(1)}, -\Gamma_{\parallel,B}^{(2)}\right)^{-1} \cdot \left(\Gamma_{\parallel,B}^{(3)}, \Gamma_{\parallel,B}^{(4)}\right), \qquad (19)$$

and $\hat{T}_{12} = \begin{pmatrix} t_{11}^F & t_{21}^F \\ t_{12}^F & t_{22}^F \end{pmatrix}, \hat{R}_{12} = \begin{pmatrix} r_{13}^F & r_{23}^F \\ r_{14}^F & r_{24}^F \end{pmatrix}, \hat{T}_{34} = \begin{pmatrix} t_{33}^F & t_{43}^F \\ t_{34}^F & t_{44}^F \end{pmatrix}, \hat{R}_{34} = \begin{pmatrix} r_{31}^F & r_{41}^F \\ r_{32}^F & r_{42}^F \end{pmatrix}$, where $r_{ab}^F, t_{ab}^F$ are Fresnel reflection and transmission coefficients into eigenmode $b$ upon incidence of eigenmode $a$. The boundary conditions Eq. (18)-(19) can be combined into a single matrix equation

$$\hat{F}_B^{-1} \cdot \hat{F}_0 = \begin{pmatrix} \hat{1} & \hat{R}_{34} \\ \hat{R}_{12} & \hat{1} \end{pmatrix} \cdot \begin{pmatrix} \hat{T}_{12} & \hat{0} \\ \hat{0} & \hat{T}_{34} \end{pmatrix}^{-1} = \begin{pmatrix} \hat{T}_{12}^{-1} & \hat{R}_{34}\hat{T}_{34}^{-1} \\ \hat{R}_{12}\hat{T}_{12}^{-1} & \hat{T}_{34}^{-1} \end{pmatrix}. \qquad (20)$$

Using Eq. (20) we can rewrite the optical response matrix $\hat{R}$ of a bianisotropic layer Eq. (15) in terms of Fresnel coefficients. First, let us express the characteristic matrix $\hat{\Delta}$ using eigenvalue decomposition (Eq. (2)): $\exp(ik_0 d\hat{\Delta}) = \hat{F}_B \cdot \exp(ik_0 d\hat{Q}) \cdot \hat{F}_B^{-1}$. From this Eq. (15) can be expressed through matrix $\hat{F}_B^{-1} \cdot \hat{F}_0$

$$\hat{R} = \left(\hat{F}_B^{-1}\hat{F}_{0+} - \exp(ik_0 d\hat{Q}) \hat{F}_B^{-1}\hat{F}_{0-}\right)^{-1} \left(\exp(ik_0 d\hat{Q}) \hat{F}_B^{-1}\hat{F}_{0+} - \hat{F}_B^{-1}\hat{F}_{0-}\right) \quad (21)$$

We introduce matrices

$$\hat{U}_+ = \hat{F}_B^{-1} \cdot \hat{F}_{0+} = \begin{pmatrix} \hat{T}_{12}^{-1} & \hat{0} \\ \hat{R}_{12}\hat{T}_{12}^{-1} & \hat{0} \end{pmatrix}, \hat{U}_- = \hat{F}_B^{-1} \cdot \hat{F}_{0-} = \begin{pmatrix} \hat{0} & \hat{R}_{34}\hat{T}_{34}^{-1} \\ \hat{0} & \hat{T}_{34}^{-1} \end{pmatrix} \quad (22)$$

Substituting Eq. (20) and (22) into Eq. (21) we express the optical response of the layer in terms of Fresnel coefficients at the boundaries of the layer in direct analogy to the Airy-Fresnel formula for isotropic layers

$$\hat{R} = \left(\hat{U}_+ - \exp(ik_0 d\hat{Q}) \cdot \hat{U}_-\right)^{-1} \left(\exp(ik_0 d\hat{Q}) \cdot \hat{U}_+ - \hat{U}_-\right) \quad (23)$$

Using Eqs. (18)-(23) we express the resonance denominators Eqs. (16)-(17) in terms of Fresnel coefficients as

$$\delta_A = 1 - r_{13}^F r_{31}^F e^{i\phi_{13}} - r_{14}^F r_{41}^F e^{i\phi_{14}} - r_{23}^F r_{32}^F e^{i\phi_{23}} - r_{24}^F r_{42}^F e^{i\phi_{24}} + e^{i\phi_{13}} e^{i\phi_{24}} (r_{13}^F r_{31}^F r_{24}^F r_{42}^F - r_{13}^F r_{32}^F r_{24}^F r_{41}^F + r_{14}^F r_{41}^F r_{23}^F r_{32}^F - r_{14}^F r_{42}^F r_{23}^F r_{31}^F), \quad (24)$$

where $\phi_{ij} = k_0 \left(q_z^{(i)} - q_z^{(j)}\right) d$.

Eq. (24) provides the physical background for resonances of bianisotropic layers and shows that they correspond to interference of 4 single roundtrip phasors $r_{ij}^F r_{ji}^F e^{i\phi_{ij}}$, which are related to propagation channels involving (a) propagation in mode *i* across the layer, (b) reflection from mode *i* into mode *j*, (c) propagation in mode *j* backwards, and (d) reflection from mode *j* into mode *i* at the original interface. Additionally, the phasors corresponding to two roundtrips across the layer participate in the interference as the last term of Eq. (23). This equation generalizes the result obtained for cavities formed by two uniaxial metasurfaces described in Eq. (7) of Ref. [130].

3. Results and Discussion

***High-k Fabry-Perot resonances in thin subwavelength bianisotropic cavities***

In Fig. 2(a) we plot the resonant denominator $\delta_A$ in logarithmic scale for the bianisotropic material illustrated in Fig. 1(a). The dependence is shown on various orientations with respect to the boundaries of a subwavelength layer with $(k_0 d) = 0.1$ [see schematic in Fig. 1(b)]. The orientations are given using Euler angles in convention $Z(\phi)Y(\theta)Z(\alpha)$, i.e. the material is first rotated around z-axis by angle $\alpha$, then around y-axis by angle $\theta$ and so on. The normal incidence $k_\parallel = 0$ is considered in Fig. 2. The orientations corresponding to the directions in which high-k modes with $k \to \infty$ propagate are given by the zeros of the *high-k characteristic function* $h(\mathbf{k}) = (\mathbf{k}^T \hat{\epsilon} \mathbf{k})(\mathbf{k}^T \hat{\mu} \mathbf{k}) - (\mathbf{k}^T \hat{X} \mathbf{k})(\mathbf{k}^T \hat{Y} \mathbf{k}) = 0$ introduced in [32,33]. The orientations of bianisotropic medium such that the $k \to \infty$ directions are perpendicular to the layer interfaces are indicated as red dashed curves in Fig. 2(a).

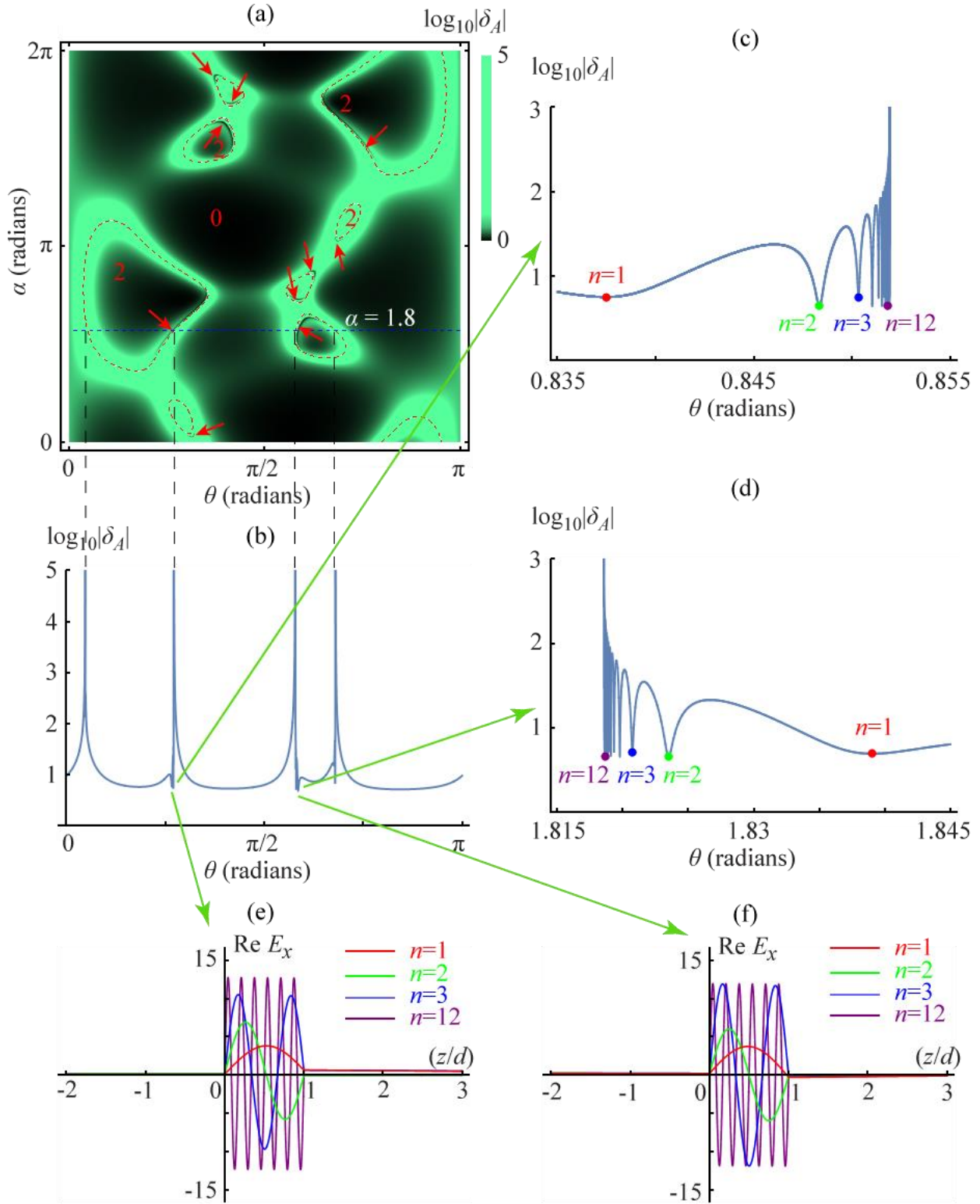

Fig. 2. High-k Fabry-Perot resonances in subwavelength bianisotropic cavity with $(k_0 d) = 0.1$ at normal incidence $\mathbf{k}_\parallel = 0$. (a) The resonant denominator $\delta_A$ as a function of the bianisotropic material orientation with respect to the layer boundaries. (b-d) cross-sections of panel (a) illustrating the high-k Fabry-Perot resonances. (e-f) Field profiles of the high-k Fabry-Perot resonances.

This curve separates the graph into regions with different number of eigenmodes with real $q_z$ which is indicated by the red numbers. For each orientation, there are 4 eigenmodes whose $q_z$ can be real or complex. In Fig. 2(a) the closed regions have 2 modes with real $q_z$ and 2 with complex. The region in the space between the closed regions has 0 modes with real $q_z$ and all 4 eigenmodes have complex $q_z$. The high-k characteristic function $h(\mathbf{k})$ is a bulk property related to the bulk Fresnel wave surface of the bianisotropic material (Fig. 1(a)). Since it serves as the separatrix between different sets of eigenmodes, the $h(\mathbf{k})$ function establishes the topological bulk-edge correspondence between the waves in the bulk bianisotropic material and the guided waves of the bianisotropic cavities and interfaces.

For the formation of Fabry-Perot resonances the eigenmodes with real $q_z$ are necessary, therefore, the Fabry-Perot resonances are only possible in the corresponding regions. In deeply subwavelength thin cavities, the high-k Fabry-Perot resonances can be supported for the orientations with high-k propagating modes close to the dashed red curves. This is seen by the dark resonance minima of $\log_{10} \delta_A$ marked by red arrows in Fig. 2(a). A cross-section of Fig. 2(a) at $\alpha = 1.8$ is shown in Fig. 2(b). One can see that zeros of $h(\mathbf{k}) = 0$ correspond to the divergences $\delta_A$; and for the orientations in the vicinity of these divergences one can find the high-k Fabry-Perot resonances, as detailed in panels 2(c) and 2(d). In Fig. 2(c) the orientations corresponding to $\alpha = 1.8$ and $\theta = 0.835 \div 0.855$ are shown. The $n = 1,2,3,12$ high-k Fabry-Perot resonances corresponding to $\lambda = nd/2$ modes are indicated by the red, green, blue, and purple dots, respectively. The field profiles for these modes are shown in panel 2(e). Note that close to the $h(\mathbf{k}) = 0$ direction, the density of the Fabry-Perot resonances diverges as $n \to \infty$. A similar situation occurs for the orientations $\alpha = 1.8$ and $\theta = 1.815 \div 1.845$, demonstrated in Figs. 2(d) and 2(f). In these panels, the high-k Fabry-Perot resonances with $n = 1,2,3,12$ are shown.

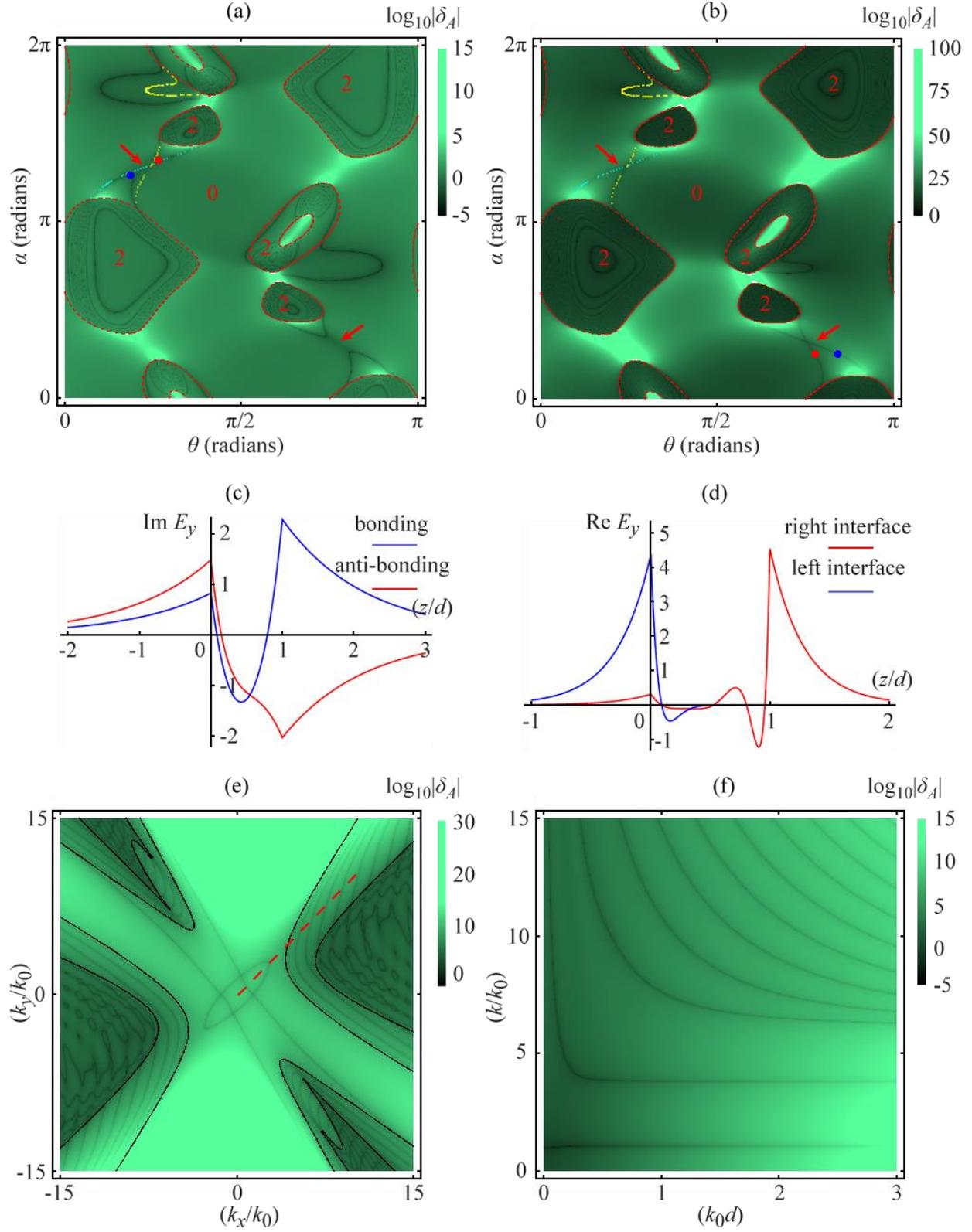

Fig. 3. Surface Electromagnetic Waves in Bianisotropic Media. (a-b) The resonant denominator $\delta_A$ as a function of the bianisotropic material orientation with respect to the layer boundaries for $q_x = 2$ and $(k_0 d) = 0.5$ in (a) and $(k_0 d) = 2$ in (b); light blue and yellow outlines correspond to SEWs at individual boundaries of the layers. Panels (c-d) – Field profiles of SEWs for $(k_0 d) = 0.5$ in (c) and $(k_0 d) = 2$ in (d) for orientations indicated by blue and red dots in panels (a) and (b). Panel (e) – the isofrequency curves of SEWs in $(k_x, k_y)$ plane for $(k_0 d) = 2$, $\theta = 1.4$, and $\alpha = 1.885$. Panel (f) – dependence of the wave number of SEWs on layer thickness propagating in the direction indicated by red dashing in (e).

One can also see the diverging density of Fabry-Perot modes with $n \to \infty$ close to the $h(\boldsymbol{k}) = 0$ direction in the subwavelength bianisotropic cavity with $(k_0 d) = 0.1$, investigated in Fig. 2.

*Surface electromagnetic waves in bianisotropic materials.*

In Fig. 3 we investigate the modes with longitudinal momentum $q_x = 2$ in thicker layers. Fig. 3(a) corresponds to $(k_0 d) = 0.5$, while Fig. 3(b) to $(k_0 d) = 2$. Due to the different longitudinal momentum the zero curve (red dashing) of the high-k characteristic function $h(\boldsymbol{k}) = 0$ separates Figs. 3(a)-(b) into somewhat different regions as compared to panel 2(a).

There are still 2 types of regions: (i) containing 2 modes with real $q_z$ and 2 with complex $q_z$ marked by the red number 2, and (ii) containing 0 modes with real $q_z$ and all 4 eigenmodes with complex $q_z$ marked by number 0. Note that due to larger thickness the layers in Fig. 3 support lower-k Fabry-Perot resonances which populate most of the regions with real $q_z$, which can be seen as dark minima curves in these regions.

Surface electromagnetic waves also appear in the region with all evanescent waves as dark minima curves. For moderately thick layers with $(k_0 d) = 0.5$ [panels 3(a) and 3(c)], these SEWs feature hybridization and their fields are characterized by bonding (blue dot in Fig. 3(a) and blue curve in Fig. 3(c)) or anti-bonding behavior (red dot and red curve respectively) depending on the phases on the interfaces of the layer. Due to the hybridization coupling of the SEW fields at different interfaces, they feature anti-crossing in Fig. 3(a) as indicated by the red arrows. Because of this, the resonance conditions for hybridized modes deviate from the orientations at which SEWs of individual boundaries propagate, as outlined with yellow and blue curves for $\theta < \frac{\pi}{2}$. These outlines are found as poles of Fresnel eqs. (18) and (19), which represent the dispersion of SEWs at bianisotropic interfaces $\det\left\{\left(\Gamma_\parallel^{(1)}, \Gamma_\parallel^{(2)}, -\Gamma_{\parallel,B}^{(3)}, -\Gamma_{\parallel,B}^{(4)}\right)\right\} = 0$, $\det\left\{\left(\Gamma_\parallel^{(3)}, \Gamma_\parallel^{(4)}, -\Gamma_{\parallel,B}^{(1)}, -\Gamma_{\parallel,B}^{(2)}\right)\right\} = 0$ [134, 105]. For $\theta > \pi/2$ the existence conditions for these single boundary SEWs are symmetrically reflected in the $\alpha < \pi$ region (not shown). In thick layers (see Fig. 3(b) and (d) for $(k_0 d) = 2$), the SEWs do not hybridize and localize at individual boundaries. The SEW resonance conditions shown in Fig. 3(b) perfectly follow the blue and yellow curves for single boundary SEWs localized at different boundaries. This can be seen from the field profiles shown in Fig. 3(d), where the blue curve corresponds to the blue dot in Fig. 3(b) and the red curve to the red dot, respectively.

In Fig. 3(e) we show an isofrequency k-curve for SEWs in the $(k_x, k_y)$ plane for $\theta = 1.4$ and $\alpha = 1.885$, in a layer with thickness $(k_0 d) = 2$. This graph features both elliptical and hyperbolic shapes of SEW isofrequency curves between the regions of guided waves in all 4 corners of this panel. We show the dependence of momentum of these modes on thickness in Fig. 3(f). This panel corresponds to the direction in the $(k_x, k_y)$ plane indicated by the red dashed line in Fig. 3(e). In Fig. 3(f) there are 2 SEW modes with k vectors depending on the thickness of the layer only for small thicknesses $(k_0 d) < 0.5$, after which there is no hybridization in these SEWs. There are also 8 propagating modes in the upper right side of this panel.

4. Conclusion

To conclude, in this paper we presented a complete theory of optical responses of arbitrary bianisotropic media and found explicit conditions for Fabry-Perot resonances and surface electromagnetic waves in such structures. We topologically related the existence conditions of these resonances to the high-k characteristic function of the bulk bianisotropic material and the taxonomy according to the hyperbolic topological classes.

5. Declarations

Availability of data and materials -- Not applicable

Competing interests – The authors declare absence of competing interests

Funding -- Not applicable

Authors' contributions -- Not applicable

Acknowledgements -- Not applicable

**Declaration of Competing Interest**

The authors declare that they have no known competing financial interests or personal relationships that could have appeared to influence the work reported in this paper.

**Acknowledgements**

We acknowledge Georgia Southern University Scholarly Pursuit Funding Award, Summer Research Session Grant, Georgia Power Student Award, College Office of Undergraduate Research (COUR) Grants from the College of Science and Mathematics at Georgia Southern University, and The Georgia Southern McNair Scholars Program.